\documentclass[preprint,11pt]{elsarticle}

\usepackage{graphicx}
\usepackage{amssymb}
\usepackage{natbib}

\journal{Physics of the Dark Universe}

\begin{document}

\begin{frontmatter}

\title{An estimate of the DM profile in the Galactic bulge region}

\author{Fabio Iocco}
\ead{fabio.iocco.astro@gmail.com}
\author{Maria Benito}
\ead{mariabenitocst@gmail.com}

\address{ICTP South American Institute for Fundamental Research, and Instituto de F\'isica Te\'orica - Universidade Estadual Paulista (UNESP), Rua Dr.~Bento Teobaldo Ferraz 271, 01140-070 S\~{a}o Paulo, SP Brazil}

\begin{abstract}
We present an analysis of the mass distribution in the region of the Galactic bulge,
which leads to constraints on the total amount and distribution of Dark Matter (DM) therein.
Our results --based on the dynamical measurement 
of the BRAVA collaboration-- are quantitatively compatible with those of a recent analysis, 
and generalised to a vaste sample of observationally inferred morphologies of the
stellar components in the region of the Galactic bulge.
By fitting the inferred DM mass to a generalised NFW profile,
we find that cores ($\gamma\lesssim$ 0.6 ) are forbidden only for very light configurations of the bulge, 
and that cusps ($\gamma\gtrsim$ 1.2) are allowed, but not necessarily preferred.
Interestingly, we find that the results for the bulge region are compatible
with those obtained with dynamical methods (based on the rotation curve) applied to outer regions of the Milky Way,  
for all morphologies adopted.
We find that the uncertainty on the shape of the stellar morphology heavily affects 
the determination of the DM distribution in the bulge region,
which is gravitationally dominated by baryons, adding up to the uncertainty
on its normalization. 
The combination of the two hinders the actual possibility to infer sound 
conclusions about the distribution of DM in the region of the Galactic bulge, and only 
future observations of the stellar census and dynamics in this region will bring us closer
to a quantitatively more definite answer.
\end{abstract}

                      
\end{frontmatter}

\section{Introduction} 
\label{sec:intro}
\par The density structure of the Dark Matter halo embedding our own Galaxy, the Milky Way (MW),
is the object of a continuous research effort, which has its roots in several decades of 
rigorous astronomical activity. 
The best known method for deriving the total mass structure of the MW is probably that
based on the use of tracers of the Rotation Curve (RC), which in turn traces the
total gravitational potential in galaxies which --like our own-- are rotation supported.
Once the total mass profile is inferred through this method, one can in principle
subtract that of the visible component (obtained through independent
observations), and fit the extracted residuals to pre-assigned spherical shapes, referred
to as DM density profiles.
Examples of the application of these method can be found in 
\cite{CatenaUllio2010}, \cite{Iocco:2011jz}, \cite{Nesti:2013uwa}, \cite{Pato:2015dua},
and a review of the history and results of such method, often referred to as ``global'', can be found in 
\cite{JustinRead2014}.
This technique does however rely on the assumption of circular orbits for the tracers of the RC, 
and an axisymmetric gravitational potential, and therefore can not be adopted in regions
where either one of the two assumptions is broken, namely at galactocentric distances
R$\lesssim$ 2.5 kpc (or, under more conservative approach R$\lesssim$ 4 kpc, \cite{2015A&A...578A..14C}).
Alternative methods, either based on a different approach like the one adopted
in \cite{Bovy:2013raa}, or in the release of the {\it ansatz} on the shape of the distribution like the one adopted
in \cite{Pato:2015tja}, also fail to address the region of the Galactic bulge.

\par Whereas one does expect the DM to be a subleading component of the
matter content in this central region of the Galaxy, our forced ignorance about it
still remains frustrating. 
In fact, the actual content and density structure of DM in such a small region (of the MW, and of other galaxies) is
at the center of a long--standing debate on whether the DM profile of galaxies is ``cored'' or ``cusped'', see e.~g. \cite{2010AdAst2010E...5D} 
which has been recently revived by the vigorous developments in numerical simulations of Galaxy formation in cosmological environments, e.~g. \cite{2010Natur.463..203G,2015MNRAS.454.2092O,2016MNRAS.456.3542T}, with the interesting finding of a DM profile which 
responds to the presence of baryons, and exhibits a core or a cusp profile in different regimes of the stellar over halo
mass, \cite{2012MNRAS.422.1231G,DiCintio:2013qxa}.
The DM profile in the GC region is also a very important ancillary quantity for ``indirect'' searches, 
which often target regions of supposedly higher DM density, like the Galactic Center,
thus making this quantity of high relevance also in the search for the very nature of the DM, e.g. \cite{Calore:2015oya,Schaller:2015mua}.

\par In order to obtain an estimate of the DM mass and density profile (and related uncertainaties) in the region of
the Galactic bulge, we follow here the rationale of the recent analysis in \cite{Hooper:2016ggc}.
We adopt an estimate of the dynamical mass within a small region around the Galactic center \cite{portail2015, Portail:2016vei},
obtained through observations of the BRAVA survey \cite{Howard:2008mu, 2012AJ....143...57K}.
Separately, we estimate the stellar mass in the same region by making use of a vast sample
of observationally inferred stellar morphologies, and fit the residuals of the two to a generalised NFW profile (gNFW).
Owing to a larger sample of stellar morphologies, and a most conservative estimate of the statistical uncertainty on
the stellar mass estimate, our results -while remaining compatible with those in  \cite{Hooper:2016ggc}- show
a much larger band of uncertainty for the slope of the profile, $\gamma$, and remain yet inconclusive.

\section{Method}
\label{sec:method}
\par We first adopt a determination of the total mass in the region of interest,
 a square box of coordinates around the Galactic Center:
\begin{center}
{\bf \([x,y,z]=[\pm 2.2,\pm 1.4, \pm 1.2]\)} ;
\end{center}
which reads
\(M_{total}=(1.84\pm 0.07)\times 10^{10}\,M_{\odot}\) .\\ 

This estimate is presented in \cite{portail2015},
after a reconstruction of the recent 3D density of red clump giants (RCG) as well as
kinematic measurements, from the BRAVA survey \cite{Howard:2008mu, 2012AJ....143...57K},
and adopted in \cite{Hooper:2016ggc} \footnote{We have explicitly tested that
our results do not change if we adopt the updated estimate of
total dynamical mass, \(M_{total}=(1.85\pm 0.05)\times 10^{10}\,M_{\odot}\),
presented by the same authors in  \cite{Portail:2016vei}.}.

It is interesting to notice that the extremes of the region in exam 
are complementary to the validity limit of analysis based on 
dynamical tracers, which as discussed are typically restricted to galactocentric radii $R > 2.5$ kpc.
In order to know the DM content, we now need information about the stellar mass in the very same region,
so as to obtain the component missing from the total gravitational budget.

\par The estimate of the stellar (baryonic) mass in the region of interest,
is where our analysis most diverges from that in \cite{Hooper:2016ggc}: 
rather than relying on the results of \cite{portail2015}, based on N-body,
made-to-measure simulations implementing a peanut-shaped bulge and
different bulge-to-disk ratio
-resulting in five different morphologies for the region of interest-,
we adopt a wider array of bulge (and disc) shapes modeled and normalized after observational data, all available in the literature.

\par This collection of three-dimensional morphologies (see Tables for references) has been first presented in 
\cite{2015NatPh..11..245I} and later adopted in \cite{Pato:2015dua,Pato:2015tja} in order to estimate the 
ignorance on the actual morphology and total mass of the baryonic component of the MW, 
comparing the existing stellar distribution profiles.
In order to obtain the baryonic mass, we integrate the three dimensional stellar density
$\rho_*^i(R,\theta,\phi)$ thus collected (where the index $i$ runs over the morphology type) over the region of interest.
At no point we do perform any average or mix the information of different bulges (nor discs): we simply present the
information available for all the morphologies in the literature in order to represent at best the current uncertainty, thus offering an estimate of the current systematics.
The normalization of the profiles is given for all of them in the source reference, however for most cases uncertainties are not available.
In order to obtain a reliable estimate on the uncertainty on the stellar mass, and at the same time to
comply with existing constraints on microlensing, we normalize the stellar mass distribution of the bulge by imposing the 
MACHO constraints on microlensing optical depth toward the region of the Galactic Center, \cite{Popowski:2004uv}.
This technique has been first described in \cite{Iocco:2011jz}, and widely adopted in the literature after then.
For the stellar component in the disc(s), we adopt a normalization based on a recent estimate of the surface stellar density \cite{Bovy:2013raa}.
We address the reader to the original references and to \cite{Pato:2015dua} for a thorough description of these techniques and test of their validity, and
we stress here that the existing observational uncertainty on the microlensing optical depth and disc stellar surface density automatically
assigns a (statistically meaningful) uncertainty to the stellar mass, which becomes thus comparable (and compared)
to that of the total mass, as inferred by the dynamical measurement of the BRAVA survey.
It is worth stressing that the bulge normalization is indeed affected 
by the stellar disc adopted: we do also account for that 
--it is the reason of the different bulge masses between the different panels of the Tables--
however the choice of the disc does not change our conclusions. 
We will however present self-consistent results for all bulges and disc morphologies in our sample.
The Galaxy also contains a disc of diffuse, interstellar gas \cite{Ferriere1998,Ferriere2007} which we neglect here,
because its contribution to the total baryonic mass in this region is below the smallest of uncertainties in our budget.

\par The procedure described above permits to derive the DM mass allowed in the region of interest, and its uncertainty,
as a simple difference between the observational derivation of the total mass $M_{tot}$, and the inference of the stellar mass from 
observationally inspired profiles $M_*^i$, where the index {\it i} runs over all the possible combination of bulges and 
discs.

For each stellar morphology in our sample we can now obtain an allowed DM mass (and its uncertainty),
simply as the difference between $M_{tot}$ and $M_*^i$ (thus the allowed DM mass will be a function of the adopted morphology),
and at this point we set out to understand what constraints does this impose on the parameters 
of a generalised NFW (gNFW) profile:

\begin{equation}
\rho_{DM} (R) = \rho_0 \left(\frac{R_0}{R}\right)^{\gamma}\left(\frac{R_s+R_0}{R_s+R}\right)^{3-\gamma} ,
\label{eq:gNFW}
\end{equation}
by identifying the regions in the ($\rho_0$, $\gamma$, $R_s$) space that would bear a mass in excess or defect with respect
to the one allowed, per each morphology.

\par It is to be noticed that --as the methodology described is self consistently defined within a small region around the Galactic Center-- 
the meaning of the index $\gamma$ is purely local, and one can see it as an effective index valid only in the region described.

\vspace{0.5cm}

\section{Results} 
\label{sec:results}

\begin{figure}[t!]
\centering 
\includegraphics[scale=0.4]{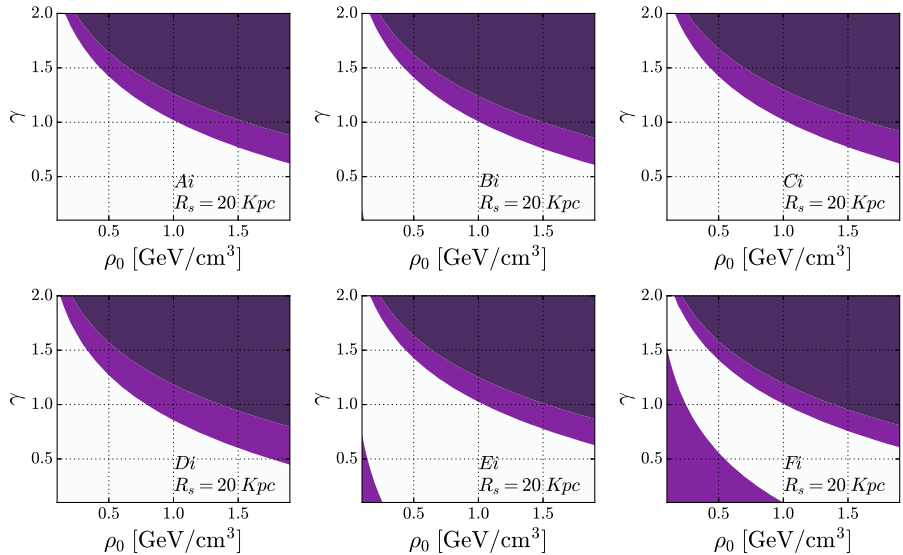}
\caption{\label{fig:bulges} Constraints in the gNFW parameter space, for an assigned value of $R_s$=20kpc. Different panels show the result of changing bulge morphologies. The disc component is fixed to \cite{Han:2003ws}.}
\end{figure}

\par
\begin{figure}[t!]
\centering 
\includegraphics[scale=0.4]{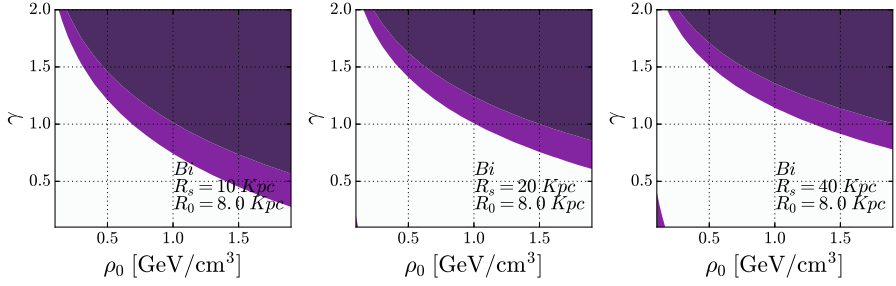}
\caption{\label{fig:rs} Constraints in the gNFW parameter space, for an assigned bulge/disc morphology, and varying the scale radius $R_s$. $R_0$=8.0 kpc.}
\end{figure}

\begin{figure}[t!]
\centering 
\includegraphics[scale=0.4]{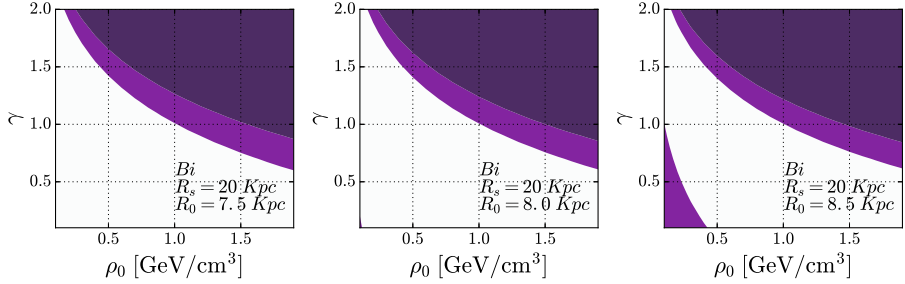}
\caption{\label{fig:r0} Constraints in the gNFW parameter space, for an assigned bulge/disc morphology, and varying the Sun's galactocentric distance $R_0$. $R_s$=20 kpc.}
\end{figure}

 Finally, we present here our results, 
displaying in all Figures the regions in the gNFW parameter space which produce
a mass compatible with that allowed by procedure described above, 
within one sigma (in white), within one and two sigmas (pale pink), 
and in tension at more than two sigmas (dark purple).
We first show the results varying the bulge morphology (which sets most of the
stellar mass in the region of interest), and by keeping constant Galactic parameters
(R$_0$, v$_0$)=(8kpc, 230km/s) and disc component.
Figure \ref{fig:bulges} displays the results for our six bulge configurations, 
in the ($\rho_0$, $\gamma$) plane, for an assigned scale radius $R_s$=20kpc.\\
The effect of the variation of scale radius is indeed small, as it can be seen in Figure \ref{fig:rs},
where we show the constraints in the same  ($\rho_0$, $\gamma$) plane for three different values of
$R_s$, spanning the extremes of the range suggested by numerical simulations; we therefore keep the choice
of $R_s$=20kpc for the following plots. We also check the effect of varying the Sun's Galactocentric distance
$R_0$ within the allowed range $R_0$ =[7.5, 8.5] kpc (notice that the technique adopted here is -unlike the RC method-
insensitive to the value of the local circular velocity $v_c$), the results can be seen in Fig \ref{fig:r0},
and motivate us to keep a constant $R_0$=8kpc for the rest of this paper.

\par As it can be easily noticed, accounting for uncertainties in the determination of the stellar mass opens
up a relatively ample region in the parameter space. While not prohibiting extremely peaked profiles, 
our analysis shows no real preference for a cuspy DM distribution in the region of interest, the lowest
values of $\gamma$ being disfavored for few configurations only.

\par
\begin{figure}[p!]
\centering
\includegraphics[width=1.1\columnwidth]{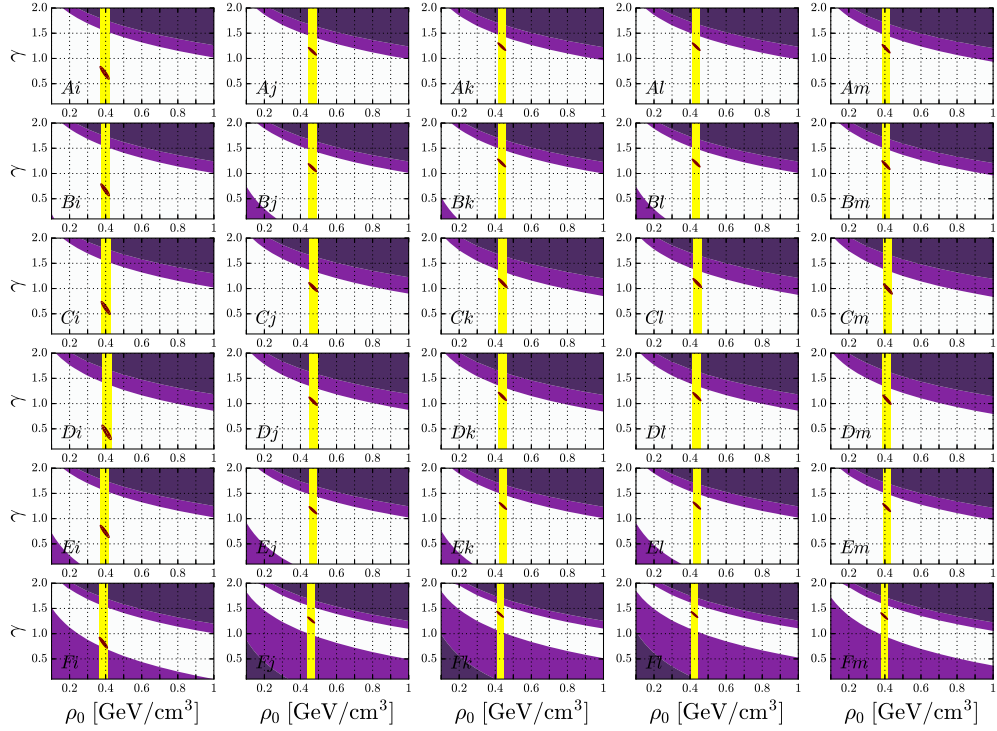}
\caption{\label{fig:complete} As in Figure \ref{fig:bulges}, for all bulge (vertical axes) and disc (horizontal axis) configurations. The small red regions are the 3$\sigma$ region as from the analysis of RC in \cite{Pato:2015dua}. The yellow bands highlight the extremes of the latter for the local DM density $\rho_0$. \(R_s=20\) kpc, \(R_0\)=8.0 kpc.}
\end{figure}

 In Figure \ref{fig:complete}, we show the effects of different discs and bulges, thus encompassing effectively all possible morphologies.

In these plots, we have also shown -for each baryonic morphology- preferred regions from the 
RC analysis in \cite{Pato:2015dua}, picking the 99.73\% ($3\sigma$) confidence region encompassed by $\chi^2 \leq \chi^2_{\textrm{bf}}+\Delta\chi^2$, where $\chi^2_{\textrm{bf}}$ 
is the $\chi^2$ of the best fit configuration and $\Delta\chi^2=11.83$ (corresponding to two fitted parameters at $3\sigma$).
While performing this comparison, one has to bear in mind that the region studied by the two analysis (which are also different in methodologies) are complementary. 
The index $\gamma$ obtained by the RC analysis is based on an comparison of the tracers of the total gravitational potential at $R>2.5$ kpc,
yet the mass enclosed at $R<2.5$ kpc weighs in the total gravitational strength, though it does not represent a leading term.

\par However, adopting the values of $\rho_0$ determined from the RC analysis can be seen as imposing a continuity of the DM profile at the boundary between the two regions. We highlight the range of $\rho_0$ allowed by the latter, in Figure \ref{fig:complete}, so that it can act as a possible prior in the determination of a favorite range of allowed $\gamma$.
We summarize the content of Figure \ref{fig:complete} in the Tables \ref{tab:values1}--\ref{tab:values5}, where we list the
stellar mass content (and its uncertainty) and the estimated DM mass, within the region of interest, for all of our bulge and disc components. 
We also report the range of $\gamma$ allowed, if one imposes a prior on the local DM density, as discussed above.

\par Interestingly, it is to be noticed that the allowed DM mass estimated with this method is compatible with zero (unsurprisingly, given that we do expect baryons to dominate the DM potential in such a small central region) for almost our entire sample of morphologies. Yet, even with the quite sizable uncertainties at hand, it is possible to put a sound upper (and in few cases, lower) limit to the estimate of the index $\gamma$,
locally for this inner region of the MW.

\section{Conclusions}\label{sec:conc}
\par We have presented an analysis of the DM density profile in the Galactic bulge region, 
following and complementing upon the recent analysis in \cite{Hooper:2016ggc}.
By adopting an observational estimate of the total dynamical mass, and a 
large array of observationally--derived morphologies for the stellar components,
we are able to estimate the current statistical uncertainties on the allowed DM mass.
We find it to vary between 6\% and 45\% the total dynamical mass. 
These figures are compatible with those of the analysis in \cite{Hooper:2016ggc}, 
yet they define a bigger range (compatible with zero at one sigma) as a consequence of
the systematic arising from our ignorance on the actual morphology of the region of the bulge,
and of the statistical on the actual normalization of the visible component.
\par Both type of uncertainties are sizable and mask away sound, general conclusions
on the preferred region of the parameter space (if the DM mass is fit to a gNFW profile):
whereas cuspy profiles are not strictly prohibited (although indexes $\gamma\gtrsim$1.5
are generally disfavored at 1 $\sigma$ at least, depending on the bulge/disc configuration), neither are
flat profiles, an index $\gamma$=0 allowed for many of the morphologies at study.
\par These results are compatible with analysis of outer regions, at galactocentric radii $R>$2.5 kpc,
performed with global techniques based on a fitting of the Rotation Curve, but this is hardly
indicative, nor surprising, given the very large degree of uncertainty described above. 
\par We argue that it is still early for a conclusive answers about the distribution of DM in the Galactic center,
and that the bottleneck of the current impasse is the estimate of the visible component morphology and normalization, 
rather than that on the total dynamical mass, in a similar fashion to what happens
for the determination of the DM profile in the outer regions of the Galaxy.

Future surveys dedicated to the stellar census and dynamics in that region, like the forthcoming Wide Field InfraRed Survey Telescope (WFIRST) will
help in the endeavour of reducing uncertainties, and move toward sounder conclusions about this fascinating problem.

\bgroup
\def\arraystretch{1.4}
\begin{table}[p!]
\center
\resizebox{\textwidth}{!}{%
\begin{tabular}{| c | c | c | c | c | c |}
\hline
Bulge &  Reference &  \(M_*^{bulge}\)  (\(\times 10^{10}\,M_{\odot}\)) &\(M_*\)  (\(\times 10^{10}\,M_{\odot}\)) & \(M_{DM}\) (\(\times 10^{9}\,M_{\odot}\)) & \(\gamma\)  \\
\hline \hline
A & \cite{Dwek:1995xu} (G2) & \(0.4 \pm 0.4\) & \(1.4\pm0.4\) & \(4 \pm 4	\) &  0.00 - 1.59 \\ \hline
B &  \cite{Dwek:1995xu} (E2) & \(0.4 \pm 0.4\)& \( 1.4 \pm	0.4 \) & \(4 \pm	4\)  &  0.00 -  1.57 \\ \hline
C & \cite{Vanhollebeke:2009ka} & \( 0.5 \pm 0.5 \) & \( 1.5 \pm	0.5 \)  & \(3 \pm 5\)   & 0.00 - 1.58 \\ \hline
D & \cite{Bissantz:2001wx} & \(0.6 \pm 0.4\) & \( 1.7 \pm	0.5 \)  &  \(2 \pm 5 \)  & 0.00 - 1.43 \\ \hline
E & \cite{zhao1996steady} & \(0.3\pm 0.4\) & \(1.4 \pm	0.4 \) &  \(4 \pm 4	\) &  0.00 - 1.59 \\ \hline
F  & \cite{robin2012stellar} & \(0.3 \pm 0.3 \) & \( 1.3 \pm	0.3\) & \(5 \pm	3\)  & 0.69 - 1.58 \\ \hline
\end{tabular}}
\caption{\label{tab:values1} Values of stellar and allowed DM mass, in the region of interest, for all bulge configurations. 
The value of index $\gamma$, as allowed at 1 $\sigma$ within the local DM density $\rho_0$ band identified
by the RC analysis (yellow line in Fig \ref{fig:complete}. Stellar disc \cite{Bovy:2013raa} (``{\it i} ''),  \(M_*^{disc} = (1.06\pm 0.11)\times 10^{10}\,M_{\odot}\);  \(R_s=20\) kpc,  \(R_0\)=8.0 kpc.}
\end{table}

\begin{table}[p!]
\center
\resizebox{\textwidth}{!}{%
\begin{tabular}{| c | c | c | c | c | c |}
\hline
Bulge &  Reference &   \(M_*^{bulge}\)  (\(\times 10^{10}\,M_{\odot}\))&\(M_*\)  (\(\times 10^{10}\,M_{\odot}\)) & \(M_{DM}\) (\(\times 10^{9}\,M_{\odot}\)) & \(\gamma\)  \\
\hline \hline
A & \cite{Dwek:1995xu} (G2) & \(0.9\pm 0.4\)&\(1.5 \pm	0.4 \) & \( 4 \pm 4\) & 0.00 - 1.46 \\ \hline
B &  \cite{Dwek:1995xu} (E2) & \(0.9 \pm 0.4\) &\(1.4 \pm	0.4 \) & \( 4 \pm 4 \) & 0.00 - 1.48 \\ \hline
C & \cite{Vanhollebeke:2009ka} & \(1.1\pm 0.5\) & \(1.7 \pm	 0.5\) & \( 2 \pm 5 \)  & 0.00 - 1.38 \\ \hline
D & \cite{Bissantz:2001wx} & \(1.1 \pm 0.4 \) & \(1.6 \pm	0.4\) &  \( 2 \pm  5\) & 0.00 - 1.35 \\ \hline
E & \cite{zhao1996steady} & \(0.8\pm 0.4\) & \(1.4 \pm	 0.4\) &  \( 5 \pm 4\) & 0.00 - 1.49 \\ \hline
F  & \cite{robin2012stellar} & \(0.6\pm 0.3\) & \(1.2 \pm	0.3\) & \( 6 \pm 3 \) & 0.96 - 1.56 \\ \hline
\end{tabular}}
\caption{\label{tab:values2} Values of stellar and allowed DM mass, in the region of interest, for all bulge configurations. 
The value of index $\gamma$, as allowed at 1 $\sigma$ within the local DM density $\rho_0$ band identified
by the RC analysis (yellow line in Fig \ref{fig:complete}. Stellar disc \cite{Han:2003ws} (``{\it j} ''), \(M_*^{disc} = (0.55\pm 0.06)\times 10^{10}\,M_{\odot}\);  \(R_s=20\) kpc,  \(R_0\)=8.0 kpc.}
\end{table}

\begin{table}[p!]
\center
\resizebox{\textwidth}{!}{%
\begin{tabular}{| c | c | c | c | c | c |}
\hline
Bulge &  Reference &  \(M_*^{bulge}\)  (\(\times 10^{10}\,M_{\odot}\)) & \(M_*\)  (\(\times 10^{10}\,M_{\odot}\)) & \(M_{DM}\) (\(\times 10^{9}\,M_{\odot}\)) & \(\gamma\)  \\
\hline \hline
A & \cite{Dwek:1995xu} (G2) & \(0.2 \pm 0.4 \) & \(1.1\pm0.4\) & \(7 \pm	4\) &  1.28 - 2.00  \\ \hline
B &  \cite{Dwek:1995xu} (E2) & \(0.2 \pm 0.4 \)& \(1.1\pm	0.4\) & \(7 \pm 4	\)  &  1.34 - 2.00 \\ \hline
C & \cite{Vanhollebeke:2009ka} & \(0.2\pm 0.5\) & \(1.2\pm	0.5\)  & \(7 \pm	5\)   & 0.96 - 2.00 \\ \hline
D & \cite{Bissantz:2001wx} & \(0.4 \pm 0.4 \) & \(1.3\pm0.5	\)  &  \(5 \pm	5\)  &  0.46 - 2.00 \\ \hline
E & \cite{zhao1996steady} & \(0.2 \pm 0.4 \) & \(1.1\pm	0.4\) &  \(7 \pm	4\) & 1.38 - 2.00\\ \hline
F  & \cite{robin2012stellar} & \(0.1 \pm 0.3 \) &\(1.1\pm	 0.3\) & \(8 \pm	3\)  &  1.55 - 2.00\\ \hline
\end{tabular}}
\caption{\label{tab:values3} Values of stellar and allowed DM mass, in the region of interest, for all bulge configurations. 
The value of index $\gamma$, as allowed at 1 $\sigma$ within the local DM density $\rho_0$ band identified
by the RC analysis (yellow line in Fig \ref{fig:complete}. Stellar disc \cite{CalchiNovatiMancini2011} (``{\it k} ''), \(M_*^{disc} = (0.93 \pm 0.10)\times 10^{10}\,M_{\odot}\);  \(R_s=20\) kpc,  \(R_0\)=8.0 kpc.}
\end{table}

\begin{table}[p!]
\center
\resizebox{\textwidth}{!}{%
\begin{tabular}{| c | c | c | c | c | c |}
\hline
Bulge &  Reference & \(M_*^{bulge}\)  (\(\times 10^{10}\,M_{\odot}\)) & \(M_*\)  (\(\times 10^{10}\,M_{\odot}\)) & \(M_{DM}\) (\(\times 10^{9}\,M_{\odot}\)) & \(\gamma\)  \\
\hline \hline
A & \cite{Dwek:1995xu} (G2) & \(1.0 \pm 0.4\) & \(1.5 \pm 0.4\) & \(4\pm	4\) &  0.00 - 1.49\\ \hline
B  &  \cite{Dwek:1995xu} (E2) & \(0.9 \pm 0.4\) & \(1.4\pm 0.4	\) & \(4 \pm 4\)  &  0.00 - 1.52 \\ \hline
C & \cite{Vanhollebeke:2009ka} & \(1.2 \pm 0.5\) & \(1.7\pm0.5	\)  & \(2 \pm	5\)   & 0.00 - 1.39 \\ \hline
D & \cite{Bissantz:2001wx} & \(1.2 \pm 0.4\) & \(1.7\pm0.4	\)  &  \(2\pm	4\)  &  0.00 - 1.38 \\ \hline
E & \cite{zhao1996steady} & \(0.9 \pm 0.4 \) & \(1.4\pm	0.4\) &  \(4 \pm	4\) &  0.00 - 1.52 \\ \hline
F  & \cite{robin2012stellar} & \(0.7 \pm 0.3\) & \(1.2\pm0.3	\) & \(7 \pm	3\)  & 1.04 - 1.61\\ \hline
\end{tabular}}
\caption{\label{tab:values4} Values of stellar and allowed DM mass, in the region of interest, for all bulge configurations. 
The value of index $\gamma$, as allowed at 1 $\sigma$ within the local DM density $\rho_0$ band identified
by the RC analysis (yellow line in Fig \ref{fig:complete}. Stellar disc  \cite{deJong2010} (``{\it l} ''), \(M_*^{disc} = (0.48 \pm 0.05)\times 10^{10}\,M_{\odot}\);  \(R_s=20\) kpc,  \(R_0\)=8.0 kpc.}
\end{table}

\begin{table}[p!]
\center
\resizebox{\textwidth}{!}{%
\begin{tabular}{| c | c | c | c | c | c |}
\hline
Bulge &  Reference & \(M_*^{bulge}\)  (\(\times 10^{10}\,M_{\odot}\)) & \(M_*\)  (\(\times 10^{10}\,M_{\odot}\)) & \(M_{DM}\) (\(\times 10^{9}\,M_{\odot}\)) & \(\gamma\)  \\
\hline \hline
A  & \cite{Dwek:1995xu} (G2) & \(0.9 \pm 0.4\) & \(1.5\pm0.4\) & \(3\pm4 \) &  0.00 - 1.49\\ \hline
B &  \cite{Dwek:1995xu}  (E2) & \(0.9 \pm  0.4 \) & \(1.5\pm	0.5\) & \(4\pm 4	\)  &  0.00 - 1.52 \\ \hline
C & \cite{Vanhollebeke:2009ka} & \(1.1 \pm 0.5 \) & \(1.7\pm	0.5\)  & \(1\pm	5\)   & 0.00 - 1.39 \\ \hline
D & \cite{Bissantz:2001wx} & \(1.1 \pm 0.4\) & \(1.7\pm	0.4\)  &  \(1\pm	5\)  &  0.00 - 1.35 \\ \hline
E & \cite{zhao1996steady} & \(0.8 \pm 0.4\) & \(1.5\pm	0.4\) &  \(4\pm	4\) & 0.00 - 1.52 \\ \hline
F  & \cite{robin2012stellar} & \(0.6 \pm 0.3\) & \(1.3\pm	0.3\) & \(6\pm	3\)  & 0.99 - 1.61 \\ \hline
\end{tabular}}
\caption{\label{tab:values5} Values of stellar and allowed DM mass, in the region of interest, for all bulge configurations. 
The value of index $\gamma$, as allowed at 1 $\sigma$ within the local DM density $\rho_0$ band identified
by the RC analysis (yellow line in Fig \ref{fig:complete}. Stellar disc \cite{Juric2008} (``{\it m} ''), \(M_*^{disc} = (0.60 \pm 0.06) \times 10^{10}\,M_{\odot}\);  \(R_s=20\) kpc,  \(R_0\)=8.0 kpc.}
\end{table}

\vspace{0.5cm}
{\it Acknowledgments}
We thank G.~Bertone, F.~Calore, M.~Pato, and P.~D.~Serpico for valuable comments on this manuscript. M.B. thanks M.~Peir\'o for valuable help on numerical recipes. \\
The authors would like to express a special thanks to the Mainz Institute for Theoretical Physics (MITP) for its hospitality and support.\\
F.~I. is supported by FAPESP JP project 2014/11070-2.

\bibliographystyle{elsarticle-num}
\bibliography{GCprofile.bib}

\end{document}